\documentclass[acmsmall,screen]{acmart}

\AtBeginDocument{%
  }

\setcopyright{acmlicensed}
\acmDOI{10.1145/3729386}
\acmYear{2025}
\acmJournal{PACMSE}
\acmVolume{2}
\acmNumber{FSE}
\acmArticle{FSE116}
\acmMonth{7}
\received{2025-02-26}
\received[accepted]{2025-04-01}

\usepackage{tcolorbox}
\usepackage{hyperref}
\usepackage{longtable}
\usepackage{tikz}
\usepackage{xspace}
\usepackage{siunitx}
\usepackage{enumerate}

\begin{document}

\title{CXXCrafter: An LLM-Based Agent for Automated C/C++ Open Source Software Building}

\author{Zhengmin Yu}
\orcid{0000-0002-0294-414X}
\affiliation{
  \institution{Fudan University}
  \city{Shanghai}
  \country{China}
}
\email{zmyu23@m.fudan.edu.cn}

\author{Yuan Zhang}
\orcid{0000-0003-0726-9996}
\affiliation{
  \institution{Fudan University}
  \city{Shanghai}
  \country{China}
}
\email{yuanxzhang@fudan.edu.cn}

\author{Ming Wen}
\orcid{0000-0001-5588-9618}
\affiliation{
  \institution{Huazhong University of Science and Technology}
  \city{Wuhan}
  \country{China}
}
\email{mwenaa@hust.edu.cn}

\author{Yinan Nie}
\affiliation{
  \institution{Fudan University}
  \city{Shanghai}
  \country{China}
}
\orcid{0009-0000-5500-7283}
\email{24210240089@m.fudan.edu.cn}

\author{Wenhui Zhang}
\orcid{0009-0006-9401-8683}
\affiliation{
  \institution{Fudan University}
  \city{Shanghai}
  \country{China}
}
\email{21009100158@stu.xidian.edu.cn}

\author{Min Yang}
\orcid{0000-0001-9714-5545}
\affiliation{
  \institution{Fudan University}
  \city{Shanghai}
  \country{China}
}
\email{m_yang@fudan.edu.cn}

\ifthenelse{\isundefined{\keepnotes}}{
\newcommand{\zhengmin}[1]{\textcolor{blue}{}}
\newcommand{\yuan}[1]{\textcolor{orange}{}}
\newcommand{\civi}[1]{\textcolor{red}{}}
\newcommand{\updated}[1]{\textcolor{black}{#1}}
}{
\newcommand{\zhengmin}[1]{\textcolor{blue}{[Zhengmin: #1]}}
\newcommand{\yuan}[1]{\textcolor{orange}{[yuan: #1]}}
\newcommand{\civi}[1]{\textcolor{purple}{[Ming: #1]}}
\newcommand{\updated}[1]{\textcolor{blue}{#1}}
}

\newcommand*\circled[1]{\tikz[baseline=(char.base)]{
            \node[shape=circle,draw=black, fill=white,text=black,inner sep=0.8pt] (char) {#1};}}

\newcommand{\name}{\texttt{CXXCrafter}\xspace}

\begin{abstract}
Project building is pivotal to support various program analysis tasks, such as generating intermediate representation code for static analysis and preparing binary code for vulnerability reproduction. However, automating the building process for C/C++ projects is a highly complex endeavor, involving tremendous technical challenges, such as intricate dependency management, diverse build systems, varied toolchains, and multifaceted error handling mechanisms. Consequently, building C/C++ projects often proves to be difficult in practice, hindering the progress of downstream applications.
Unfortunately, research on facilitating the building of C/C++ projects remains to be inadequate.
The emergence of Large Language Models (LLMs) offers promising solutions to automated software building. Trained on extensive corpora, LLMs can help unify diverse build systems through their comprehension capabilities and address complex errors by leveraging tacit knowledge storage. Moreover, LLM-based agents can be systematically designed to dynamically interact with the environment, effectively managing dynamic building issues.
Motivated by these opportunities, we first conduct an empirical study to systematically analyze the current challenges in the C/C++ project building process. Particularly, we observe that most popular C/C++ projects encounter an average of five errors when relying solely on the default build systems. Based on our study, we develop an automated build system called \name to specifically address the above-mentioned challenges, such as dependency resolution. Our evaluation on open-source software demonstrates that \name achieves a success rate of 78\%  in project building. 
\updated{Specifically, among the Top100 dataset, 72 projects are built successfully by both \name and manual efforts, 3 by \name only, and 14 manually only}.
Despite the slightly lower performance, \name can save tremendous manual efforts and can also be easily applied to a wider range of applications automatically. 
\end{abstract}



\maketitle

\section{Introduction}

Open Source Software (OSS) is extensively used across various sectors and plays a critical role in powering modern technological systems, from critical infrastructure to innovative applications.
\updated{Many foundational software systems, such as operating systems and database management systems, are \updated{written in} C/C++ \cite{Cimportant}.}
Vulnerabilities in these systems can cause significant damage~\cite{Thornburg01}(e.g., the Heartbleed bug in OpenSSL~\cite{Heartbleed}), making automated vulnerability detection for C/C++ OSS essential.
\updated{Additionally, automated analysis of C/C++ OSS is crucial for areas like vulnerability management, quality assurance, and performance optimization.}

Building C/C++ \updated{software}~\cite{wiki:softwarebuild} involves compiling, resolving dependencies, linking libraries, configuring environments, and managing platform-specific challenges.
\updated{These processes are critical for performing automated program analysis, especially for dynamic analysis. Such tasks require the project to be built into a binary beforehand.} 
However, a significant gap remains in the realm of automated analysis of C/C++ OSS \cite{gygi2021cppbuild}: the absence of a standardized, automated method for building repositories from C/C++ source code. Bridging this gap is essential for enhancing the efficiency and effectiveness of program analyses.
The significance of automatically building software from source code for automated analysis can be summarized from two key aspects:
\updated{(1) \textbf{Facilitation of Static Program Analysis}:}
Many static analysis tasks rely on intermediate representations (IR), such as LLVM IR \cite{llvm}. This \updated{typically} requires that the project successfully installs dependencies and can be compiled.
\updated{(2) \textbf{Enablement of Dynamic Program Analysis}:}
Dynamic program analysis \updated{like} fuzzing, also requires that the program \updated{be compilable} from source code, \updated{particularly when} source-level instrumentation is needed. Automatically built projects can also assist with several downstream tasks, such as automated vulnerability reproduction tasks.

Existing research efforts~\cite{hassan2017automatic, horton2019dockerizeme, 10.1145/3510003.3510132, 10.1145/3551349.3556923, 10.1145/3597503.3639209, 8453189} mainly focus on Java/Python, while C/C++ remains underexplored due to its higher complexity.
Unlike the relatively unified and automated build and package management tools in Java (e.g., \textit{Maven}~\cite{apache_maven}, \textit{Gradle}~\cite{gradle}) and JavaScript (e.g., \textit{NPM}~\cite{npm_docs}), or Python’s convenient \textit{pip}~\cite{pip}, the C/C++ ecosystem contains over 20 distinct build systems~\cite{awesome_cpp}, with lower levels of standardization and automation, posing significant challenges.

To better understand the automation of C/C++ project builds, we investigate the build systems of 100 popular open-source C/C++ projects across 10 different categories, using their default build commands \updated{and settings}. The study shows that more than 70\% of these projects fail to be built successfully without manual intervention, \updated{suggesting} that most C/C++ projects require additional configuration, such as downloading dependencies or setting compilation parameters. To further investigate the root causes of these failures, we manually fix the errors encountered during the build process \updated{guided by the failure messages iteratively until successful completion}. In total, we encounter 384 errors across 79 projects, and spend over 153 man-hours to resolve them. This \updated{underscores} the \updated{significant} challenges in automating the C/C++ build process.

\textbf{Challenges.}
Drawing from the root causes and insights gathered in our study, we summarize the following challenges associated with C/C++ build automation: 

\begin{itemize}
    \item \textbf{Challenge~1: Complexity of Dependency Management.}
    C/C++ projects often rely on \updated{substantial} external libraries and tools, \updated{which require careful} management and configuration of dependencies. \updated{Although package management tools like \textit{Conan} \cite{conan} and \textit{vcpkg} \cite{vcpkg} are available, they support different sets of libraries and have distinct usage patterns, which makes dependency management a complex task.}
    \item \textbf{Challenge~2: Diversity of Build Systems and Compilation Options.}
    C/C++ projects adopt at least 20 different build systems (such as \textit{Makefile} \cite{gnu-make}, \textit{CMake} \cite{cmake}, \textit{Autotools} \cite{gnu-automake-autotools}, and \textit{SCons} \cite{scons}), each with unique syntax and configuration requirements. Additionally, these projects employ a wide array of compilers (such as \textit{GCC} \cite{gcc} and \textit{Clang} \cite{clang}) and toolchains, each with its own options and configuration methods. 
    Such diverse build systems and toolchains would trigger substantial errors when building large-scale real projects. 
    \item \textbf{Challenge~3: Complexity of Error Diagnosis and Debugging.}
    The diverse build process in C/C++ projects often generate many error messages at multiple levels, such as pre-processing, compilation, and linking, which often vary greatly across different projects. 
\end{itemize}

\textbf{Our System.}
Large Language Models (LLMs) are renowned for their strong capabilities in understanding complex documentations \cite{hopkins2023classifyingcomplexdocumentscomparing, zou2024docbenchbenchmarkevaluatingllmbased}, generating structured instructions \cite{wallace2024instructionhierarchytrainingllms, jiang2024sketchtoolkitstreamliningllm}, and resolving errors \cite{grotov2024untanglingknotsleveragingllm, tyen2024llmsreasoningerrorscorrect}.
Inspired by such abilities, we investigate whether LLMs can extend their effectiveness to the domain of build systems and error resolution.
In particular, we apply LLMs (\textit{i.e.,}~\textit{GPT\-4o}) to solve specific build issues identified in our empirical study and observe that it can successfully address several of them (as demonstrated in the experiment in Section \ref{sec: overall}).
For instance, during \textit{GCC} builds, the LLM can automatically install dependencies like \textit{GMP}, \textit{MPFR}, \textit{MPC}, and \textit{Flex}, and prompts 64-bit compilation, thus avoiding errors from the default build instructions and simplifying dependency management. 
Such results demonstrate LLM's potential capability in this domain.
At the same time, it also indicates that for complex project builds, which involve multi-step processes, the effectiveness of standalone LLMs is limited.
Relying solely on LLMs can only address a small fraction of errors, highlighting the need for more refined strategies capable of continuously addressing build failures.

To address the above challenges, we propose an LLM-based agent system named \name that leverages LLMs to dynamically manage complex build processes. The system consists of three modules: the \textbf{Parser Module}, the \textbf{Generator Module}, and the \textbf{\updated{Executor} Module}. 
Specifically, 
the \textbf{Parser Module} automatically extracts and parses relevant information from the repositories, such as dependencies and build system configurations. The \textbf{Generator Module} utilizes LLMs to generate candidate build solutions \updated{(i.e., Dockerfiles, which include shell scripts for the entire software build process)} based on the parsed information. Additionally, the Generator is responsible for modifying the candidate build solutions in response to error feedback from the Executor. The \textbf{\updated{Executor} Module} oversees the build process in the Docker container where the build is performed, capturing error messages and determining whether the build is successful. The Generator and Executor form a dynamic interaction loop, continuously addressing build issues until the process completes successfully.
\updated{Our design effectively addresses the three challenges as mentioned above.}
In particular, the Parser can identify the required dependencies to avoid potential dependency errors. Besides, \name also employs an automated, iterative feedback process powered by LLMs to dynamically identify and install dependencies, thus effectively addressing issues such as uncertain dependencies or version conflicts (Challenge 1).
Furthermore, \name leverages LLMs' rich domain knowledge via nested prompt templates to unify different build systems and compilation options (Challenge 2).
For the Challenge 3, \name captures real-time feedback during the build process, enabling efficient error diagnosis and debugging by adapting to both known and new errors \updated{arising} during \updated{the} build.

We evaluate \name on both the aforementioned 100 popular C/C++ projects and the larger Awesome-CPP dataset \cite{awesome_cpp}, which includes 652 projects across various categories. 
Specifically, \name successfully builds 587 out of the 752 projects, achieving a success rate of 78\%. \updated{This} significantly outperforms other heuristic approaches (39.01\%) and bare LLM (34.22\%). Though its overall performance does not surpass the build success rate achieved by humans, \name resolves three projects that cannot be successfully built through human efforts. Our analysis for these three projects \updated{shows} that \name leverages the implicit build knowledge embedded in the LLM and the powerful retrieval capabilities of its parser module, offering unique advantages even compared to human efforts in project builds. Additionally, a component analysis demonstrates its effectiveness in designing agents capable of handling complex tasks. \updated{Finally, we assess the efficiency and cost, and the results show that it takes 875 seconds together with a financial cost of \$0.41 per successful build}. These evaluation experiments underscore the practical value of \name.

\textbf{Contributions.} This paper makes the following main contributions:
\begin{itemize}
    \item \textbf{Originality:} To our best knowledge, we are the first to explore the idea of utilizing LLM agent to automate C/C++ build process, and our study demonstrates promising results. 
    \item \textbf{Empirical Study:} We conduct an empirical study on the build processes of 100 popular open-source C/C++ projects to understand the current state of build tools. By identifying and categorizing 384 build errors, we provide a comprehensive analysis of the challenges to automate C/C++ builds, offering key findings to the root causes of such failures.
    \item \textbf{Approach:} We propose \name, an LLM-based agent system designed to automate the build process for large-scale C/C++ repositories. In particular, \name dynamically manages dependencies, resolves build issues, and diagnoses errors, effectively addressing the challenges such as handling various build systems and installing dependencies.
    \item \textbf{Evaluation:} Through extensive evaluations on 752 projects, \name achieves an impressive build success rate of 78\%, demonstrating its pioneering effectiveness in C/C++ build automation. Our research has the potential to support downstream program analysis efforts.
\end{itemize}

\section{Background \& Related Work}
\textbf{Software Building.}
Software building \cite{wiki:softwarebuild} converts code into executables or libraries, involving tasks like dependency resolution, compilation, and linking. For \updated{large} projects, automated build systems become essential, as manual handling becomes impractical. These systems streamline the \updated{process}, managing tasks efficiently. Different programming languages have specific build systems: Java uses \textit{Apache Ant} \cite{apache_ant}, \textit{Maven} \cite{apache_maven}, and \textit{Gradle}, while JavaScript relies on \textit{NPM} \cite{npm_docs}, and Python uses \textit{setuptools} \cite{python_setuptools}. In C/C++ projects, tools like \textit{CMake}, \textit{Make}, \textit{Ninja}, and \textit{Bazel} are \updated{frequently} used. Additionally, building differs from compiling, which is just one part of the broader building process, and Continuous Integration (CI), where building is a prerequisite for integration.

\updated{Several studies focus on automating software builds, mostly for languages like Java, with fewer addressing C/C++ projects.}
Hassan et al. \cite{hassan2017automatic} investigate Java build failures, revealing that 86 out of 200 projects fail to build automatically using default commands. Other studies have explored build \cite{10.1145/3510003.3510132, 10.1145/3551349.3556923, 10.1145/3597503.3639209, 8453189} and CI failures \cite{10.1145/3338906.3338917}. For \updated{example}, Lou et al. \cite{10.1145/3368089.3409760} analyzed 1,080 build issues from Stack Overflow related to \textit{Maven}, \textit{Ant}, and \textit{Gradle}, finding that 67.96\% of the issues were resolved by modifying build scripts for plugins and dependencies. Similarly, Olivier et al. \cite{10.1145/3688842} analyzed over 1.2 million build logs from Google’s \textit{OSS-Fuzz} service to identify common failure patterns.
In the context of C/C++, we only found \textit{CPPBuild} \cite{gygi2021cppbuild} for automating the build process. But it is limited to \textit{CMake}, \textit{Make}, and \textit{Autotools}, resulting in lower accuracy for open-source projects with other build systems. Furthermore, while some works focus on containerization techniques and Dockerfile generation \cite{10.1145/3238147.3240470, 10.1145/3645092, rosa2023automaticallygeneratingdockerfilesdeep}, they typically do not \updated{address} building software from source.

\textbf{LLMs and Agents.}
LLMs have shown outstanding performance across multiple dimensions, including semantic understanding \cite{tang2023largelanguagemodelsincontext}, code generation \cite{electronics10243150}, and implicit knowledge storage \cite{zhang2024distillingimplicitmultimodalknowledge}.
However, they still face several limitations~\cite{liu2023agentbench, ibm2023aiagents}, such as solving complex tasks, maintaining context over long interactions, executing actions in real-world environments, and engaging in dynamic, multi-turn dialogues. \updated{LLM-based agents, designed to address these challenges, integrate more advanced functionalities.} They are \updated{increasingly} used in a variety of scenarios~\cite{barua2024exploring}, including code generation \cite{opendevin2024, githubcopilot} and security tasks~\cite{deng2024pentestgpt}, showing significant promise for future advancements.

\begin{table}[htbp]
\caption{The Top 100 Projects and Their Categories}
\label{tab:3-1}
\begin{center}
\resizebox{\textwidth}{!}{
\begin{tabular}{ll}
\toprule
\textbf{Field} & \textbf{Projects} \\
\midrule
Kernel and System Programming Tools & GCC, 8CC, Codon, Mold, Clang, LLVM-project, Gdb, Linux, Reactos, RT-Thread\\

Game Development & Godot, GamePlay, raylib, DOOM, Entt, Stockfish, Cocos2d-x, Rpcs3, OpenRCT2, Minetest \\

Image Processing & Guetzli, Openalpr, Openpose, Aseprite, Tesseract, Mozjpeg, Libfacedetection, Libjpeg-turbo, Libjxl, Simd \\

AI Frameworks and Tools & Tensorflow, Mediapipe, LocalAI, XGBoost, OpenCV, GPT4All, Llama.cpp, Caffe, Paddle, Mxnet \\

Database Development & Rethinkdb, Mongo, Leveldb, Rocksdb, Sqlitebrowser, Duckdb, Foundationdb, Arangodb, Scylladb, Kvrocks \\

Video Processing & Srs, FFmpeg, Libvpx, Openh264, Vireo, Theora, X265, Blender, Shotcut, Libde265 \\

Audio Systems & Sonic-pi, OpenFrameworks, Rnnoise, Soloud, Aubio, Libsndfile, MuseScore, Wav2letter, Libsoundio, AudioFlux \\

Web Frameworks & Treefrog-framework, Civetweb, Pistache, Cpprestsdk, Crow, Drogon, Facil.io, Lwan, Oatpp, Userver \\

GUI & NanoGUI, RmlUi, Elements, Libui, Xtd, Flameshot, Qv2ray, Polybar, DearPyGui, Webui \\

IDEs & Geany, Code::Blocks, Jucipp, Color\_coded, CodeLite, rtags, serenity, qt-creator, cquery, Irony-mode \\
\bottomrule
\end{tabular}
}
\end{center}
\end{table}

\section{Empirical Study}
\label{sec:empirical}
In this section, we conduct an empirical study to assess the current status of building C/C++ projects, aiming to determine how effectively existing build systems can handle the complexities and challenges of real-world projects.
\updated{Our research team consists of 4 programmers, each with extensive experience in C/C++ development and building.}
\updated{Specifically}, we manually attempt to build 100 widely-used C/C++ projects, devoting approximately 153 man-hours to resolving the generated build failures. Out of the 100 projects, 86 have been built successfully, while the remaining projects either require excessive time budget or encounter unresolved issues.  Additionally, we analyze the errors encountered during the building process and summarize the root causes. The research questions, datasets, and study results are presented in detail as follows.

\textbf{Research Questions.} Referring to a recent study \cite{hassan2017automatic}, which investigates the build mechanism and ecosystem of Java, we design the following research questions for the study on C/C++ OSS:
\begin{itemize}
\item \updated{\textbf{RQ1 (Default Build Success Rate):}} What proportion of popular C/C++ projects can be successfully built using their respective build systems and default build commands?
\item \updated{\textbf{RQ2 (Build Failure Causes):}} What are the major root causes of the observed build failures among these projects?
\end{itemize}
\begin{figure*}[htbp]
\centerline{\includegraphics[width=0.9\textwidth]{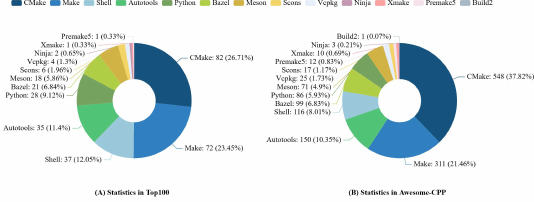}}
\caption{\updated{The Statistics of Build Tools used in the Top 100 and Awesome-CPP Datasets~(introduced in Section \ref{sec:evaluation}).}}
\label{fig: chart}
\end{figure*}

\textbf{Dataset.}
For our empirical study, we construct a dataset \updated{(hereinafter referred to as Top100)} via selecting the top 100 most popular open-source C/C++ projects from GitHub, spanning 10 distinct categories to ensure diversity and comprehensiveness. These categories include foundational projects such as operating systems, database management systems, as well as emerging projects like AI frameworks. The projects, as summarized in Table \ref{tab:3-1}, are mostly the top 10 in their respective fields based on star ratings, except for those that do not meet the following requirements. Since our builds are conducted on a Linux system, we exclude any projects that are incompatible with Linux builds (e.g., \textit{CnC\_Remastered\_Collection}). Additionally, repositories that are not fully open-source (e.g., \textit{AppCode}, \textit{Cvelop}) or do not qualify as complete projects (e.g., \textit{3d-game-shaders-for-beginners}, \textit{minimp3}) are also excluded. We focus on these repositories because they are frequently analyzed and studied in downstream applications such as program analysis, making them ideal candidates for our research. Additionally, as popular projects, they exemplify common practices and challenges in building C/C++ projects within the open-source community.

\begin{table}[h]
  \caption{Results of Executing Default Build Commands on the Top100 Dataset}
  \label{tab:default}
  \resizebox{0.9\textwidth}{!}{
  \begin{tabular}{lcl}
    \toprule
    \textbf{Category} & \textbf{\#Success} & \textbf{Successfully Built Projects}\\
    \midrule
    Kernel and System Programming Tools & 2 &  8CC (Make), Mold (CMake)\\
    Game Development & 1 &  Entt (CMake)\\
    Image Processing & 5 &  Guetzli (Bazel), Libfacedetection (CMake), Libjpeg-turbo (CMake), Libjxl (CMake), Simd (CMake)\\
    AI Frameworks and Tools & 3 &  XGboost (CMake), OpenCV (CMake), Llama.cpp (CMake)\\
    Database Development & 3 &  Leveldb (CMake), Duckdb (CMake), Kvrocks (CMake)\\
    Video Processing & 1 &  X265 (CMake)\\
    Audio Systems & 2 &  Libsndfile (CMake), Libsoundio (CMake)\\
    Web Application Frameworks & 3 &  Civetweb (Make), Facil.io (CMake), Oatpp (CMake)\\
    GUI & 1 &  Webui (CMake)\\
    IDE & 0 &  None\\
    \midrule
    \textbf{Sum} & 21 &  N/A\\
  \bottomrule
\end{tabular}
}
\end{table}
\subsection{Success Rate of Default Build Commands (RQ1)}
\label{sec:rq1}
To answer \updated{\textbf{RQ1}}, we \updated{employ} a three-phase process to apply default build commands to each C/C++ project. 
\updated{In the \textbf{first phase}}, we gather the most commonly used build commands for popular build tools through an extensive review of online tutorials and documentation. For example, we \updated{choose} `\textit{make}' for Makefile-based projects, `\textit{mkdir build \&\& cmake .. \&\& make}' for \textit{CMake} combined with \textit{make}, and `\textit{./configure \&\& make}' for \textit{Autotools}.
\updated{A} complete \updated{list} of build systems and their corresponding default commands \updated{is provided} in the appendix \cite{zenodo_cxxcrafter}.
\updated{In the \textbf{second phase}}, we select the build systems used by the 100 projects. We manually inspect the project’s source code directory to identify the build system. For projects that support multiple build systems, we determine the primary system and entry files \updated{(the files used to initiate the build process)} based on the official documentation. If the documentation does not offer a clear recommendation, we randomly select one to proceed with. In cases where the selected build system fails in the subsequent steps, we switch to another one. If the chosen system succeeds, the process completes.
In the \updated{\textbf{third phase}}, we apply the appropriate build commands to each project. To ensure consistency, all builds are executed separately within a newly installed \textit{Ubuntu 22.04} Docker environment, without any pre-installed dependencies. If a project has specific OS requirements, we switch to the required system.

During the build process, we document the build systems used in the 100 projects, as shown in Figure \ref{fig: chart}. The statistics reveal significant variability in the build systems employed by popular projects. In particular, most projects support \textit{CMake} and \textit{Make}, with these two systems often being used in combination. 
The results of applying the default build commands to the Top100 dataset are presented in Table \ref{tab:default}. As shown in the table, only 21 projects are successfully built, highlighting that even well-known and actively maintained projects demonstrate low compatibility with default configurations. 
For the remaining 79 projects, we observe the failure reasons can be attributed to a lack of specific setups, which can be mainly categorized into 3 types. 
\updated{\textbf{First}}, 51 projects encounter dependency-related errors, where required dependencies, such as \textit{libpng} when building \textit{mozjpeg}, are missing and not automatically installed.
\updated{
For projects with missing dependencies, we manually review the project's documentation, including files like ``README'', ``Contribution'', ``Compile'', and ``Building'', to check for any information on dependencies required before building. Out of the 51 projects, 28 have missing dependencies that are not mentioned in their documentation. Many projects do not clearly specify which dependencies are required, forcing developers to spend extra time addressing these issues.
}
\updated{\textbf{Second}}, 17 projects face issues related to incompatible build system versions or missing tools. For example, the \textit{bazel} version required for building \textit{mediapipe} does not meet the requirements. 
\updated{\textbf{Third}}, 11 projects fail due to incorrect build commands, such as needing to specify the target as `\textit{build}' when running `\textit{make}' for \textit{LocalAI}.
\updated{In total, resolving these issues for the 79 failed projects requires additional, non-default configurations across all three categories.}

\begin{tcolorbox}[colback=gray!10, colframe=black, boxrule=0.5mm, before skip=5pt, after skip=5pt, boxsep=0mm, left=1mm, right=1mm, top=1mm, bottom=1mm]
\textbf{Finding1:} The build systems of C/C++ projects vary significantly, yet the level of automation among existing systems remains relatively low. Furthermore, many projects often require additional specific setup steps to build successfully.
\end{tcolorbox}

\begin{table}[h]
\caption{Results of the Build Process by Humans on the Top100 Dataset}
\begin{center}
 \resizebox{0.6\textwidth}{!}{
\begin{tabular}{lcl}
\toprule
\textbf{Category} & \textbf{\#Success} & \textbf{Not Successfully Built Projects}\\
\midrule
Kernel and System Programming Tools & 8 & Codon, RT-Thread\\
Game Development & 8 & DOOM, Cocos2d-x\\
Image Processing & 10 & None\\
AI Frameworks and Tools & 8 & LocalAI, Paddle\\
Database Development & 7 & FoundationDB, ArangoDB, Scylladb\\
Video Processing & 10& None\\
Audio Systems & 7 & OpenFramework, MuseScore, Wav2letter\\
Web Frameworks & 10& None\\
GUI & 9 & Qv2ray\\
IDEs & 9 & Cquery\\
\midrule
\textbf{Sum} & 86 & N/A\\
\bottomrule
\end{tabular}
}

\label{tab3}
\end{center}
\end{table}

\subsection{Root Causes of Build Failures in Actual Build Processes (RQ2)}
To answer \updated{\textbf{RQ2}}, we continue building the 79 C/C++ projects that initially failed \updated{with} the default build commands \updated{by} systematically investigating each build failure. Leveraging expert knowledge and online resources, 4 \updated{programmers} resolve errors one by one, documenting each issue and verifying the resolution by ensuring the error no longer \updated{occurs}. 
In total, we have successfully built 65 out of the 79 projects. However, 14 projects \updated{cannot} be built for two reasons: \textit{unresolved source code errors} and \textit{exceeding the four-hour build time limit}.
\updated{Among these 14 projects, 9 encounter errors that cannot be resolved (e.g., the ``Unknown \textit{CMake} command `\textit{harfbuzz\_Populate}' '' in \textit{MuseScore}). We determine that these errors are unlikely to be fixed because similar issues have been reported by other developers in the official GitHub repositories, yet the project maintainers have not provided effective solutions. By searching the official GitHub issues using keywords from the error messages, we find that, of the 9 issues, 7 are still open and 2 are closed. However, even for the closed issues, the proposed solutions do not resolve our build problems.
Additionally, 5 out of 14 projects fail due to timeouts. Based on our observation, projects that exceed 4 hours typically do not resolve independently. Compiling large projects, such as the \textit{Linux kernel}, takes less than 20 minutes on our server, and the four-hour window allows sufficient attempts to address any issues. Therefore, we consider projects that exceed this time limit as failures to avoid unnecessary time expenditure.
}

After completing all the build processes, we \updated{resolve} a total of 384 errors, nearly 5 errors per project on average. By conducting a systematic taxonomy, we \updated{categorize} the root causes of these failures, as summarized in Table~\ref{tab:issue-categories}. The build failures of C/C++ projects are classified into three main categories: library issues, build toolchain issues, and configuration issues. In addition to these, we also identified other factors that contributed to the failures, such as code errors within the projects, which are classified as other issues. The following introduces the details.

\begin{table}[h]
  \centering
  \caption{Root Causes of Build Errors in the Building of the Top 100 Projects by Humans}
  \label{tab:issue-categories}
  \resizebox{0.9\textwidth}{!}{
  \begin{tabular}{llcp{10cm}}
    \toprule
    \textbf{Category} & \textbf{Subcategory} & \textbf{Count} & \textbf{Description} \\
    \midrule
     & Library Not Installed & 263 & Required libraries are not installed, resulting in missing dependencies during the build. \\
     \textbf{Library Issues} & Library Not in Path   & 10  & Libraries exist but are not included in the build system's paths such as `\textit{LD\_LIBRARY\_PATH}', so the system is not able to locate them. \\
                            & Library Version Inconsistency & 11 & The installed library version is inconsistent with what the project requires. \\
    \cmidrule(lr){2-4}
    &\textbf{Sum}            &284 & \\
    \midrule
    & Build System Version Conflict & 6 & The build system, such as \textit{CMake} or \textit{Make}, is incompatible or outdated, causing build failures. \\
    \textbf{Build Toolchain Issues}                                 & Other Tools Missing or Conflicting & 58 & Build tools, excluding the build system, such as compilers and package managers like \textit{pkg\-config} and \textit{vcpkg}, are either missing or have version conflicts. \\
    \cmidrule(lr){2-4}
    &\textbf{Sum}                    &64& \\ 
    \midrule
    & OS/Platform Incompatibility & 7 & Incompatibilities arising from the operating system or hardware architecture, including discrepancies in system versions or hardware support for libraries such as \textit{CUDA} and \textit{cuDNN}.\\
             \textbf{Configuration Issues}                                      & Incorrect Build Commands & 10 & Non-standard or incorrect build commands, including missing flags or options. \\
                                                  & Project Configuration Issues & 13 & The lack of project-specific configurations, such as a defined file structure, initialization scripts, or dependency installation scripts. \\
    \cmidrule(lr){2-4}
    &\textbf{Sum}                                  &30& \\
    \midrule
     & Memory Issues & 1 &Errors due to insufficient memory or memory allocation problems during the build process. \\
     \textbf{Other Issues}                     & Source Code Issues & 3 & Errors in the source code, such as syntax errors, undefined references, or incorrect logic. \\
                          & Unstable Builds in Certain Branches & 2 & Instability in certain project branches, resulting in inconsistent build outcomes. \\
    \cmidrule(lr){2-4}
    &\textbf{Sum}          &6&\\
    \midrule
    \textbf{Total} & &  384 & \\
    \bottomrule
  \end{tabular}
  }
\end{table}
\subsubsection{Library Issues}
Library issues often occur when the required libraries are either not installed, not placed in system environment paths, or have incompatible versions. These issues typically result in errors such as ``library not found'' or ``undefined reference'' as the compiler or linker is unable to resolve the symbols or functions defined in those libraries. 
Among open-source projects, developers often share only the core source code and exclude the installed libraries to keep the repository concise. However, this can lead to library-related errors when others attempt to build the project without the necessary dependencies installed.
This issue occurs a total of 284 times in our study, making it the most frequent problem encountered during C/C++ project builds. Compared to other build systems, such as \textit{Maven} or \textit{Gradle} in Java \cite{hassan2017automatic}, we find that C/C++ build systems generally make less effort to automatically reinstall removed libraries. To some extent, this may be due to the more complex nature of C/C++ dependencies and the lack of a unified package management tool like those found in higher-level languages such as Java or Python. These issues can be further categorized into three sub-categories as follows.

\textbf{Library Not Installed.}
In our empirical study, most library issues are attributed to missing libraries, with 263 out of 284 cases falling into this category. These missing libraries are typically, though not always, removed from open-source projects by developers to save space or for other reasons. As a result, builders manually download them through methods such as using package managers or building from source. For instance, during the build of \textit{libde265}, \textit{OpenRCT2}, and \textit{minetest}, \textit{SDL2} is not found and needs to be installed using \textit{apt}.

Errors related to missing libraries frequently occur during the preparation phase when build systems checking for dependencies. However, if left unresolved, they can also surface later during the compilation or linking phases, as observed in the building processes of projects like \textit{rpcs3}, \textit{aseprite}, and \textit{mxnet}. While package management tools like vcpkg and Conan exist for C/C++ development, they are not as widely adopted or standardized as those used in higher-level languages like Java.

\textbf{Library Not in Path.}
This issue arises when libraries are installed but not included in the system’s search paths, such as `\textit{LD\_LIBRARY\_PATH}', \updated{preventing} the build system \updated{from locating} them. In our study, this occurs 10 times, causing errors during compilation or linking phase when \updated{dependencies cannot be resolved}. For example, \textit{MuseScore} fails to build because the file `\textit{FindQt6Qml.cmake}' is not found in `\textit{CMAKE\_MODULE\_PATH}'.

\textbf{Library Version \updated{Inconsistency}.}
This issue occurs when the installed library version is inconsistent with what the project requires. Due to API or behavioral discrepancies, this leads to incompatibilities during the dependency management, linking, or compilation phases. In our study, this issue is observed 11 times in projects such as \textit{Shotcut}, \textit{OpenPose}, and \textit{Sonic-Pi}, where these conflicts resulted in build failures. Resolving such issues typically involves updating the project to accommodate the installed library version or reverting to an older, compatible version.

\subsubsection{Build Toolchain Issues}
Build toolchain issues refer to problems related to missing or incompatible versions of tools necessary for the build process, such as compilers, linkers, or other essential utilities. These issues typically arise when the project's toolchain is not fully specified or when the available version does not meet the project's requirements. In our study, this occurs 64 times. These toolchain issues can be further divided into two categories, as outlined below.

\textbf{Build System Version Conflict.}
This sub-category error occurs 6 times in our study. We ensure that the corresponding build systems are installed by default, thus avoiding missing toolchain issues. However, version mismatches occasionally occur. For example, in the \textit{wav2letter} project, the environment requires a minimum \textit{CMake} version of 3.29.2, but the version available in the default \textit{APT} repositories is 3.25.1. Due to the unique role of build systems within the toolchain, we classify this type of error as a separate sub-category.

\textbf{Other External Tools Missing or \updated{Conflicting}.}
The toolchain also includes external utilities such as debuggers, linker and profilers, which may be necessary for certain stages of the build or testing process. Incompatible or missing versions of these tools caused issues in 58 cases. For example, missing or conflicting versions of utilities like \textit{GDB} or \textit{Valgrind} can lead to failures during debugging or performance analysis stages.

\begin{tcolorbox}[colback=gray!10, colframe=black, boxsep=0mm, left=1mm, right=1mm, top=1mm, bottom=1mm]
\textbf{Finding 2:} Library issues (e.g., library not installed, version inconsistency) are the most significant challenges in C/C++ project building, followed by build toolchain issues and configuration issues. 
\end{tcolorbox}

\subsubsection{Configuration Issues}
\label{configure_issues}
Configuration issues occur when a project's build scripts are misconfigured or incompatible with the specific environment. These issues include platform or operating system incompatibilities, incorrect build options, and misconfigured files.

\textbf{System or Equipment Incompatibility.}
Certain projects are designed to run exclusively on specific operating systems or hardware platforms, and attempting to build them on an unsupported platform often results in failures. For example, projects like \textit{OpenPose} recommend using Ubuntu versions between 14 and 20, while older projects such as \textit{OpenAPLR} suggest \textit{Ubuntu 16.04}. Additionally, hardware-specific requirements, such as the absence of a GPU, can prevent the building of projects reliant on \textit{CUDA} and \textit{cuDNN}. In our evaluation, such errors occurred 7 times.

\textbf{Incorrect Build Commands}
Build instructions often require specific setups, such as configuring environment variables, cross-compilation, or managing dependencies. For example, when building for a different architecture like ARM, a toolchain file must be specified to ensure proper compilation: `\textit{cmake -DCMAKE\_TOOLCHAIN\_FILE=path/to/arm\_toolchain.cmake ..}'. Without such configurations, the build process may fail or produce incorrect results.
 
\textbf{Project Configuration Issues}
This error occurred 13 times and is typically caused by missing project-specific configurations, such as hardcoded paths or dependencies hosted on private sources. For example, the \textit{gameplay} project requires files (e.g., \textit{gameplay-deps}) from a specific URL. Without performing these required custom setups, the build process is bound to fail.

While we also encountered issues such as source code errors and unstable versions, which occurred 6 times. Our study focuses primarily on build system-related problems. We have documented these issues as they pose significant barriers to successful builds.

\begin{tcolorbox}[colback=gray!10, colframe=black, before skip=5pt, after skip=5pt, boxsep=0mm, left=1mm, right=1mm, top=1mm, bottom=1mm]
\textbf{Finding 3:} Build errors in C/C++ projects can occur at various stages, including dependency resolution, compilation, linking, or runtime setup. These issues are diverse in nature, as they vary depending on the build tools and project characteristics involved at each stage.
\end{tcolorbox}

Building C/C++ projects is challenging, even for human developers, and automating this process adds further complexity. Based on our empirical study, we summarize the key challenges in automating C/C++ builds.
Dependency management is a frequent challenge during dynamic builds, involving the identification, downloading, and resolution of issues. While Software Composition Analysis (SCA) studies \cite{woo2021centrisprecisescalableapproach, 10.1145/3597503.3639209, 10172788} address dependency issues, they fail to detect non-local third-party libraries (TPLs) before building. Research like \textit{CCScanner} \cite{tang2022towards} examines package management tools, but build-specific issues like alias conflicts and version mismatches remain unaddressed. Static analysis is insufficient for dependencies that are conditional, dynamically loaded, or tied to build environments with varying compiler flags and OS requirements. Additionally, dependencies often originate from multiple sources, such as package managers (e.g., \textit{apt}) or source code, and the obstacles in downloading them further complicate the resolution process.
Second, in our study, we manually write extensive shell scripts and perform debugging within Dockerfiles, utilizing various tools like build systems, compilers, and package managers. The diversity of these tools and their commands makes it difficult to standardize the build process with fixed rules, presenting a major challenge in automation.
Lastly, build errors can occur at any stage, including preprocessing, compilation, linking, or even due to external factors like network issues or hardware limitations, such as insufficient memory, can also cause build failures. These errors vary significantly, making it difficult to apply generalized error-handling strategies. Furthermore, the solutions to these problems are often scattered across various sources, requiring extensive expertise or the ability to conduct in-depth research through documentation, community forums, and other resources.
All these challenges hinder the automation of building C/C++ projects.

\section{Methodology}
In light of the challenges discussed in Section \ref{sec:empirical}, we design an agent \name to streamline the building of C/C++ projects by leveraging LLMs to handle various stages of the building process.
\subsection{Overview}
Our approach is driven by the broad capabilities of LLMs across multiple dimensions, including semantic understanding \cite{tang2023largelanguagemodelsincontext}, code generation~\cite{electronics10243150}, and implicit knowledge storage \cite{zhang2024distillingimplicitmultimodalknowledge}. Existing studies show that LLMs' semantic understanding enables the performance of code analysis tasks~\cite{fang2024largelanguagemodelscode} and assists in bug and error comprehension~\cite{jin2023inferfix, lee2024unified}, demonstrating potential for interpreting diverse error messages in the build process. Their robust code generation capabilities allow developers to create applications in various programming languages~\cite{zheng2024surveylargelanguagemodels}, with promising potential for automatically generating build instructions and bash scripts to resolve build errors. Additionally, through training on extensive corpora, LLMs implicitly store vast amounts of knowledge across multiple domains, helping to address issues in various fields~\cite{rawat2024diversitymedqaassessingdemographicbiases, lu2021codexgluemachinelearningbenchmark, talmor2019commonsenseqaquestionansweringchallenge}. LLMs may have been trained on large-scale open-source resources, including GitHub issues and Stack Overflow~\cite{carlini2021extracting, mukherjee2024mediumlmscodeera, kaddour2023challenges}, which contain numerous build-related problems and solutions, further underscoring their potential in tackling challenges related to software construction.

\updated{However, the effectiveness of directly using LLMs for building is limited. As shown in the experimental results in Section~\ref{sec:evaluation}, using bare LLMs with prompts successfully generates build solutions for only about 30\% of the projects.}
This is because the build process for many C/C++ projects involves multi-faceted errors, including those arising from different stages of the build. Relying solely on the LLM can only address a small fraction of such errors. An iterative approach is needed to continuously resolve issues as they arise. To address this, we propose an LLM-based agent that dynamically manages the build process through iterative feedback mechanisms. This agent autonomously resolves errors in real-time, adjusting and refining build decisions based on evolving conditions. The framework not only reduces the need for manual intervention but also enhances build reliability and success rates.

\begin{figure*}[htbp]
\centerline{\includegraphics[width=0.8\textwidth]{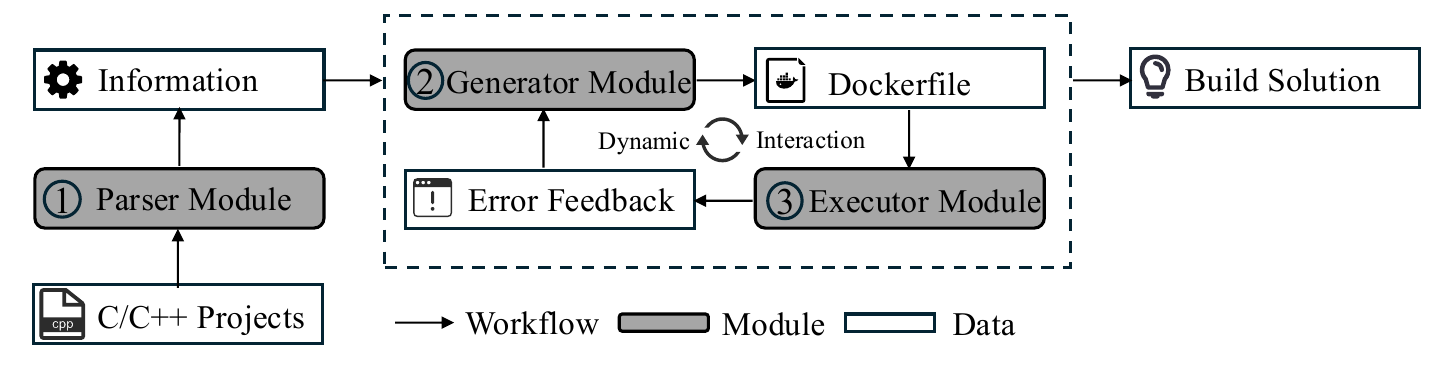}}
\caption{The Overall Framework of \name}
\label{fig: overview}
\end{figure*}

As illustrated in Figure \ref{fig: overview}, the \name is comprised of three essential modules:
\begin{itemize}
\item \textbf{Parser Module}: \updated{This module} automatically extracts and analyzes key build-related information from the \updated{project directory}, \updated{encompassing} dependencies, environment \updated{settings}, and relevant documentation that facilitate the build process. This ensures that all essential data is available for the subsequent stages of the workflow. Additionally, we leverage the LLM's semantic understanding capabilities to overcome two key obstacles: identifying the valid build system entry file and retrieving helpful documentation.
\item \textbf{Generator Module}: \updated{This module} utilizes LLMs to generate a Dockerfile \updated{that includes build procedure code} based on the parsed information, ensuring that necessary dependencies, environment settings, and configurations are correctly specified. The module also modifies the Dockerfile in response to error feedback from the \updated{Executor} Module, ensuring an adaptive approach to resolving build issues.
\item \textbf{Executor Module}: \updated{This module} oversees the build process in containers by execute Dockerfile, providing a consistent and clean build environment for testing whether the build solution succeeds. Specifically, it captures errors and logs, feeding them back to the \updated{Generator} Module, forming a dynamic \updated{interaction} loop that continuously addresses errors until completion.
\end{itemize}

\name uses five types of prompts for different use cases, incorporating techniques such as RAG and nested prompt templates, as detailed in Section \ref{sec: prompt}.

\begin{figure*}[htbp]
\centerline{\includegraphics[width=\textwidth]{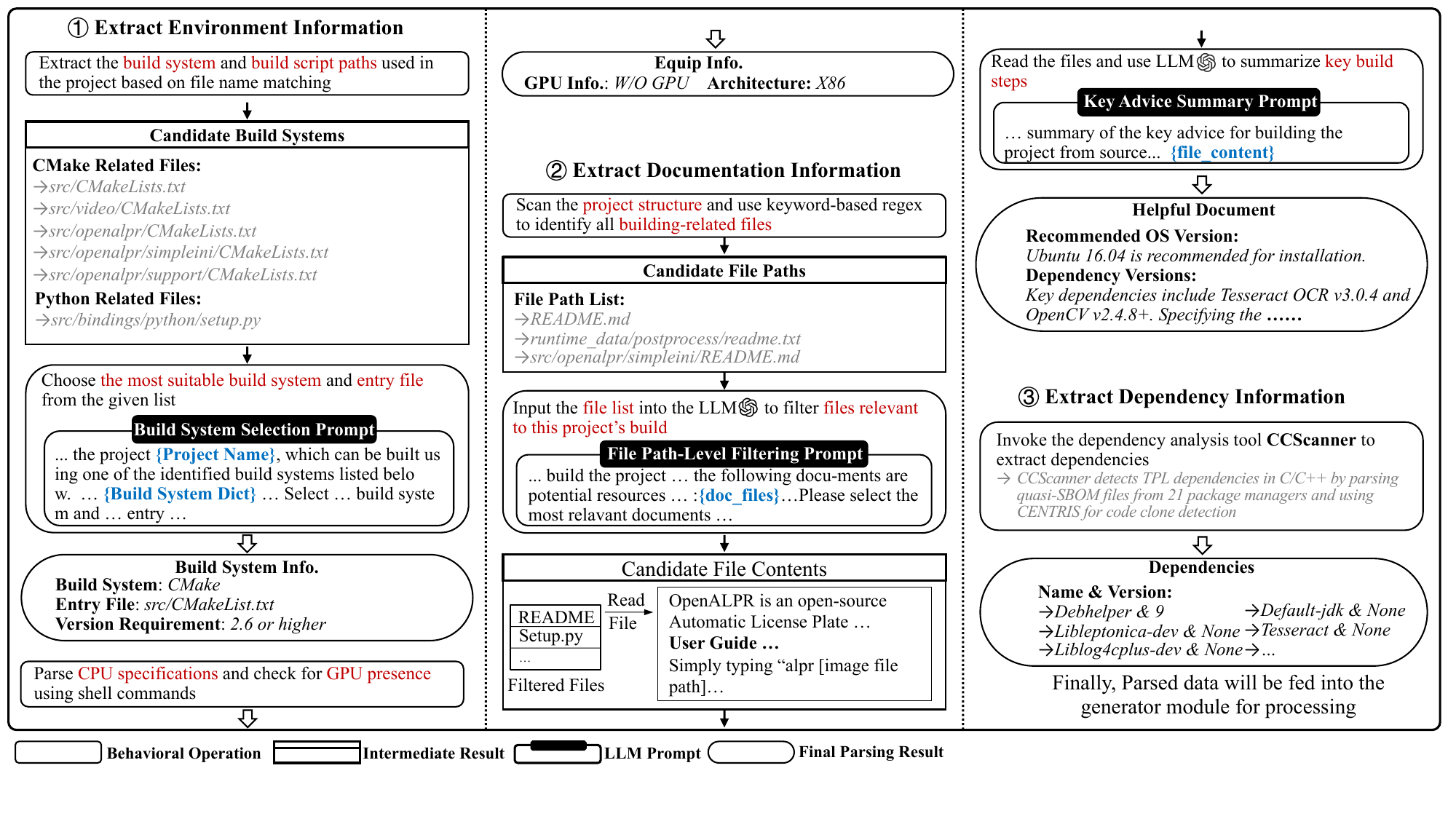}}
\caption{\updated{The Parser Module. It is responsible for automatically extracting and analyzing information such as dependencies, build system information, which is crucial for subsequent build processes.}}
\label{fig: parser}
\end{figure*}

\subsection{Parser Module}
\label{parser}
The \updated{Parser} Module analyzes local projects \updated{to} extract critical information for \updated{the} software's \updated{environment preparation} and compilation. 

It employs three specialized extractors \updated{(see Figure \ref{fig: parser})} to gather data, including environment settings, dependency details, and helpful build documentation.
\updated{\circled{1} In the \textbf{Environment Information Extractor}, \name uses basic shell commands like `\textit{lscpu}' and `\textit{uname -a}' to capture system details, including CPU specs, OS, and their versions. This information is crucial for addressing issues discussed in Section \ref{configure_issues}, such as installing software for specific architectures or ensuring the correct GPU/CPU driver versions. It plays a key role in ensuring compatibility and optimization.}
\updated{\circled{2} The \textbf{Dependency Information Extractor} takes the entire source code folder as input and outputs the names and versions of all required dependencies, helping to prevent conflicts and ensure software stability.
Existing research on dependency identification falls into two categories: some studies \cite{woo2021centrisprecisescalableapproach, 10172788, 10.1145/3597503.3639209} use Software Composition Analysis (SCA), but SCA can't recognize third-party libraries (TPLs) before build time, as many TPLs are not available locally at that stage. \textit{CCScanner} \cite{tang2022towards} detects TPL dependencies in C/C++ by parsing quasi-SBOM files from 21 package managers and using \textit{CENTRIS} \cite{woo2021centrisprecisescalableapproach} for code clone detection.
We use it in our parser module to statically extract dependency names and versions.
It is worth noting that while the statically extracted dependencies help address library-related issues, they do not fully resolve them. Specifically, they are incapable of handling dynamic errors such as aliasing (mismatched resolved and downloaded library names) or version conflicts, which only manifest during the execution process. Static dependency analysis at this stage is insufficient, necessitating the use of the generator and executor modules for dynamic resolution.}
\updated{\circled{3}} The \updated{\textbf{Useful Documentation Extractor}} collects relevant build instructions and configuration guides, aiding \name in troubleshooting and understanding the build process. \updated{As shown in Part 2 of Figure \ref{fig: parser}, it scans the source code folder and applies two rounds of filtering. First, it uses keyword-based regular expressions to identify build-related files and remove irrelevant ones. Then, it performs finer filtering using LLMs, based on the project name and document path, to exclude unrelated files. Finally, it reads the filtered files and uses the LLM to summarize key build information, ultimately obtaining the relevant documents for the build process.}

The parser module faces 2 key obstacles: identifying the correct build system and entry file, as well as retrieving useful documentation for the build process.
First, many projects employ multiple build systems, each with several build files. Expert knowledge is required to determine which build system and entry file are suited to compile the entire project. We address this by leveraging LLMs combined with tailored prompts. For example, in the \textit{OpenALPR} project (Figure \ref{fig: parser} Part 1), both \textit{CMake} and Python are present, but the LLM correctly identifies \textit{CMake}, recognizing the Python paths as interface files rather than the main project.
Second, some projects include useful documentation that aids the build process, but traditional rule-based methods struggle to locate this information. To address this, we develop a RAG system to search for relevant content. For example, in Figure \ref{fig: parser}, we retrieved documentation from the ``README.md'' file, which recommended installing \textit{Ubuntu 16.04} and provided advice on dependency versions to help avoid potential compatibility issues.

\subsection{Generator Module}
The generator module is responsible for creating and modifying build solutions. \updated{In \name, build solutions are defined using Dockerfiles, enabling the construction of C/C++ software in clean and reproducible environments. While Shell or Python scripts can also be used, Docker often offers higher flexibility and consistency. Its ability to generate clean system images ensures that the resulting Dockerfiles can be executed reliably across different environments.}

Upon receiving the output from the parser, the generator produces an initial version of the Dockerfile. We have designed curated Embedded Prompt Templates (detailed in Section \ref{sec: prompt}), \updated{which provide structured guidance to the LLM by embedding predefined formats and placeholders within the prompts.}
These templates ensure the Dockerfile creation process is structured and consistent.
The generator begins modifying the Dockerfile when the executor encounters a failure, utilizing the error message and the recently executed Dockerfile. We retain all modification history within the same session of the LLM and prioritize clearing the oldest resolved issues when the context limit is reached, allowing the model to reference recent decisions during the modification process.
Our key methodology in the generator module is the design of Embedded Prompt Templates. These templates offer structured guidance to the LLM by embedding predefined formats and placeholders within the prompts. Drawing from the building experience in Section \ref{sec:empirical}, we have systematically outlined the structure of Dockerfiles in the prompt, encompassing essential components such as system and tool installation, package management updates, dependency installation, project-specific configurations, and build-related instructions. This structured approach ensures consistency and adherence to best practices, promoting the generation of standardized yet flexible build solutions.

\subsection{Executor Module}
\label{sec: exec}
\updated{The executor module is responsible for executing the Dockerfile generated by the generator module. It monitors the entire build process to detect errors. During the building process, the executor tracks the executed commands and logs detailed traces. If the build fails, the executor sends the error messages back to the generator. This initiates an optimization process, creating a dynamic interaction loop between the generator and executor. This loop continues until a successful build solution is achieved or the maximum number of iterations is reached. Additionally, the executor implements an LLM-based discriminator on the build instructions and logs. This ensures the success of the build and helps identify and resolve errors comprehensively.}

\updated{
A critical challenge in designing the executor module is accurately verifying whether the project has been successfully built. We employ the Python Docker SDK to capture the execution results within the Docker container and save these as log files. However, certain build instruction errors may lead to issues that Docker cannot detect. One such scenario arises when a project implements custom error handling, which may suppress the generation of error messages. For example, in \textit{LocalAI}, the \textit{Makefile} includes error handling for build targets, meaning that even if the wrong target is selected, Docker will not report any build errors. Another issue occurs when the Dockerfile generated by the LLM lacks essential build instructions (e.g., `\textit{make}'). In this case, while no errors may be reported, no actual building operation takes place. We refer to these situations as ``non-error failures''. Due to the diverse nature of the outputs in these cases, traditional rule-based or keyword-matching error detection methods often fail to reliably identify such build failures.}

\updated{
To address the challenge, we design an LLM-based discriminator to identify these build failures. In designing the LLM discriminator, we incorporate two key insights from our manual construction process:
Static criterion: The Dockerfile should include build and compile instructions (e.g., `\textit{make}', `\textit{cmake --build}'), and the build target must match the default or primary components as described in the project's documentation.
Dynamic criterion: We store log files (an example is available in our project \cite{zenodo_cxxcrafter}) generated during the build process. By analyzing these logs, we can confirm whether the build commands are executed successfully. Logs from successful builds typically show compile progress (e.g., `\textit{[ 3\%] Building CXX object..}') and test progress (e.g., `\textit{Performing Test C\_FLAG\_WALL \- Success}').
The discriminator’s judgment process is divided into two steps.
First, we design prompts to guide the LLM in making judgments based on these two key criteria. Second, to further mitigate hallucinations, we introduce a reflection mechanism to re-validate the ``judgment process'' of the first step. If the ``judgment process'' did not strictly adhere to the two criteria, the build is deemed a failure, thus minimizing FP.
}
\updated{
When providing information to the discriminator, the executor carefully controls the context length and selects the log segments most relevant to state determination. In the case of ``error-type failures'', which are typically direct and concise, the executor inputs the most recent 50 execution logs into the LLM for accurate error detection and analysis. When no errors are reported, and since determining ``non-error failures'' often requires more contextual information, the executor inputs the Dockerfile and the last 200 lines of the log. If the input exceeds the LLM's context length limit, a sliding window mechanism is used, prioritizing the retention of the most recent logs to ensure effective resolution of new errors.}

\updated{
To evaluate the effectiveness of the LLM-based discriminator, we examine the accuracy of 4 LLMs—\textit{DeepSeek-v2}, \textit{DeepSeek-v3}, \textit{GPT-4o}, and \textit{GPT-4o~mini}—on the Top100 dataset. We manually check and validate the discriminator's judgments during the build process. Out of the 400 build processes, 249 are classified as successful. We manually verify these 249 samples and find all judgments to be correct. Regardless of the LLM used, the discriminator accurately identifies all successful builds. This validates the effectiveness of the LLM-based discriminator design.}

\subsection{Prompt Design in \name}
\label{sec: prompt}
We have designed a set of prompts tailored to 5 specific scenarios, including build system and entry point identification, RAG for documentation parsing, initial Dockerfile creation, Dockerfile modification, and build success discriminator.
These prompts, developed based on expert knowledge and refined through iterative experimentation, incorporate strategies such as nested prompt templates and RAG to address task complexity. The complete set of prompts is provided in the appendix file\cite{zenodo_cxxcrafter}.
In the design process, several challenges arise when prompting LLMs to effectively complete building tasks.  \textbf{Challenge 1} involves breaking down complex problems when generating build solutions. We address this by using embedded prompt templates to dynamically inject parsing information, such as in our Dockerfile generation prompt (Section 2.3 in Appendix file \cite{zenodo_cxxcrafter}), which dynamically fills in parsed data. Additionally, we provide the LLM with strategic guidance in the form of requirement notes. \textbf{Challenge 2} stems from unclear project-specific build processes and details. To resolve this, we utilize RAG to retrieve relevant files from the project’s source code directory, such as documentation RAG and build system identification prompts (Section 2.1 and 2.2 in Appendix file\cite{zenodo_cxxcrafter}). Finally, \textbf{Challenge 3}, related to token limitations, arises during Dockerfile modification. To effectively manage error feedback, we retain error messages and decisions within a single session to ensure continuity. However, when the context exceeds the token limit, we remove the earliest resolved issues to maintain focus on the current task.

\begin{figure*}[htbp]
\centerline{\includegraphics[width=0.8\textwidth]{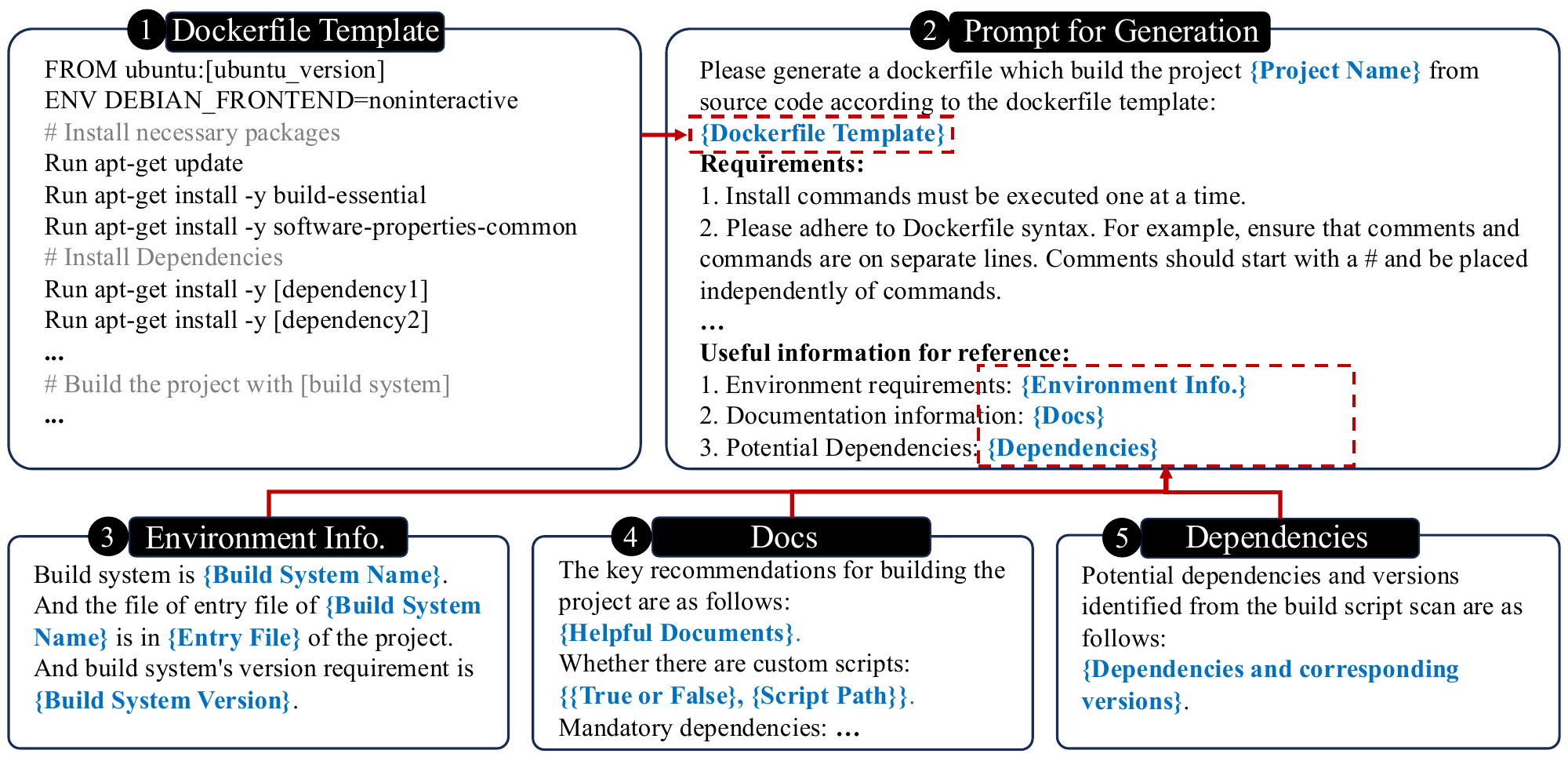}}
\caption{\updated{Prompt of the \updated{Generator} Module. \updated{The prompts corresponding to numbers 3, 4, and 5 are from the Parser’s results, with blue text indicating dynamically parsed content.}}}
\label{fig: prompt}
\end{figure*}

\updated{
\textbf{An Example of Prompt.}
As shown in Figure \ref{fig: prompt}, this prompt is used to generate a Dockerfile. Specifically, we combine information obtained from the parser with pre-defined templates for built-in prompts to create the final prompt that generates the Dockerfile. The prompts corresponding to numbers 3, 4, and 5 in the figure include the information parsed by the parser. 
Prompt 1 is a Dockerfile template that guides the LLM to structure the Dockerfile correctly, breaking down steps such as basic environment setup and dependency installation. In addition, this prompt includes specific requirements, such as correctly handling line breaks in comments, to ensure that the generated Dockerfile is free from syntax errors.
}

\section{Evaluation}
\label{sec:evaluation}
We implement \name in Python, without relying on LLM frameworks \updated{like} \textit{Langchain}.
Our implementation ensures a clear modular structure and strong scalability, enabling easy upgrades and replacements of components and tools.
\name consists of 1,664 lines of code and uses 5 different types of prompts. 
In the experiments, \name uses \textit{GPT-4o} as the default LLM, with the dynamic interaction limit set to 10 by default.
The execution environment for our build solution is managed through \updated{Python} Docker \updated{SDK}. Our experiments are conducted on three \textit{Ubuntu 22.04} servers with varying hardware configurations. The first machine is equipped with two Intel Xeon 6330 processors, 512 GB of RAM, and 3 TB of HDD storage. The second and third machines feature four Intel Xeon 8260 processors, 256 GB of RAM, and 3.37 TB of HDD storage each.

\textbf{Research Questions.}
Our evaluation aims to address the following research questions:
\begin{itemize}
\item \textbf{RQ3 (Effectiveness)}: How many C/C++ projects can be automatically built by \name?
\item \textbf{RQ4 (Ablation Study)}: How does each component within \name contribute to the overall build performance?
\item \textbf{RQ5 (Case Study)}: How does \name resolve build issues that manual methods fail to address, and what specific advantages does it offer in handling complex C/C++ projects?
\item \textbf{RQ6 (Efficiency and Cost)}: What is the efficiency and cost of using CXXCrafter?
\end{itemize}

 \textbf{Dataset.}
\updated{Two datasets are used for evaluation.}
\updated{The first dataset, Top100, is described in Section \ref{sec:empirical}. The second dataset, from Awesome-CPP \cite{awesome_cpp}, includes a broader collection with 58.6K stars as of September 2024. It covers a wide range of C++ libraries, frameworks, and tools, providing a comprehensive testbed for evaluating \name’s performance across diverse real-world C/C++ projects.}
\updated{To ensure there are no duplicates between datasets and that all projects are buildable}, we remove any \updated{overlapping projects with the Top100 dataset} and manually exclude non-C/C++ projects based on the criteria outlined in Section \ref{sec:empirical}. After filtering, 652 distinct projects remain for evaluation. For all projects, we use the latest available version for experimentation.

\updated{
\textbf{LLMs Selection.}
We select four LLMs: \textit{GPT-4o}, a high-performance closed-source model; \textit{GPT-4o~mini}, a more affordable alternative of \textit{GPT-4o}; \textit{DeepSeek-v2} with 236B parameters and \textit{DeepSeek-v3} with 671B parameters, both open-source models that excel in code-related tasks.
}

\updated{
\textbf{Baselines.}
We select 3 types of baselines.
(1) \textbf{Default Build Commands}: We have collected over 20 common C/C++ build systems and their associated instructions (see Appendix~\cite{zenodo_cxxcrafter}). Based on this collection, we develop an automated script to execute default or commonly used build commands. The script first identifies potential configuration files, such as \textit{Makefile} or \textit{CMakeLists.txt}. It then identifies all possible build systems from the configuration files and executes their corresponding build instructions in sequence.
(2) \textbf{Programmers}: The manual building methods used in Section~\ref{sec:rq1}.
(3) \textbf{Different Bare LLMs}: We also explore the performance of different bare LLMs. These models use the same prompts as the \name generator but lack the information provided by \name's parser. Additionally, there is no dynamic iterative process if the build fails.
}

\updated{
\textbf{Metrics of Success.}
We determine the success of builds by manually inspecting the Dockerfile instructions and the corresponding execution outputs.
During this inspection, we follow two criteria to efficiently assess success as follows:
(1) Static Criterion: The Dockerfile must contain the necessary build-related instructions, and the build target should align with the primary components as specified in the project documentation.
(2) Dynamic Criterion: We analyze the execution logs generated during the building process to ensure that the build commands are executed properly and that the process completes without errors.
Only projects that satisfy both criteria are considered as successful builds. We further evaluate these metrics (see Section \ref{effectiveness_metrics}), confirming that builds meeting these criteria yield outputs consistent with those produced by manual builds and demonstrate correct functionality. These criteria are the same as those in the executor (see Section \ref{sec: exec}), with the key difference being that we perform manual checks to prevent misjudgments by LLMs.
}

\subsection{Overall Effectiveness Evaluation (RQ3)}
\label{sec: overall}
To address \textbf{RQ3}, experiments are conducted on two datasets. 
For \name, \updated{the default dynamic interaction step limit is set to 10, with \textit{GPT-4o} serving as the core LLM due to its superior performance in trials.} We also evaluate the build performance of \name using another powerful open-source model, \textit{DeepSeek-v3}, while keeping all other settings the same. Additionally, we compare the results with those of the Default Build Commands and the bare LLMs, as mentioned above.
\updated{For all build results, we manually inspect and verify their correctness.}

As shown in Table \ref{tab:overall_results}, \name demonstrates significant superiority. \updated{For the Top100 dataset}, \name \texttt{(Default)} successfully builds 75 projects, significantly surpassing other methods. The Default Build Commands tool achieves 21 builds, while the bare LLM models, show a similar performance, with 23 and 17 successful builds, respectively. In the Awesome-CPP collection, \name achieves 512 successful builds, greatly outperforming Default Build Commands (272 builds) and the bare LLMs (\updated{264 for \textit{DeepSeek-v3}} and 215 for \textit{GPT-4o}).

The Default Build Commands approach achieves a 39.01\% success rate. While this method proves reliable for simpler projects, it struggles with more complex or non-standard build configurations, resulting in a relatively low success rate. The bare LLMs (\textit{DeepSeek}, \textit{GPT-4o}, and \textit{GPT-4o~mini}) demonstrate even lower success rates of 38.43\%, 31.65\%, and 19.81\%, respectively. These findings suggest that while LLMs have some capacity to handle build tasks, their effectiveness remains limited without further domain-specific optimization. In some cases, they perform worse than rule-based methodologies. Notably, \textit{GPT-4o~mini}, with a 19.81\% success rate, exhibits significant limitations when applying a smaller LLM to complex build processes.

In stark contrast, \name achieves a 78.10\% success rate, showing a marked improvement over all other methods. This outcome underscores the effectiveness of \name’s modular design, which allows it to adapt efficiently to diverse build scenarios. The substantial gap between \name and the other methodologies emphasizes the importance of specialized agent in automating complex tasks like C/C++ project builds. 

\updated{Overall, \name significantly outperforms both the bare LLMs and the heuristic build tool, demonstrating high success rates and the potential to reduce the time and efforts required for large-scale OSS building, making it a valuable tool in modern development workflows.}

\begin{tcolorbox}[colback=gray!10, colframe=black, boxrule=0.5mm, boxsep=0mm, left=1mm, right=1mm, top=1mm, bottom=1mm]
\textbf{Finding 4:} Without a carefully designed iterative framework, LLMs remain inadequate for addressing the inherent complexity and multi-stage processes of project building.
\end{tcolorbox}

\begin{table}[htbp]
\caption{Experimental Results Between \name and Baselines.}
\label{tab:overall_results}
\normalsize
\resizebox{0.7\textwidth}{!}
{
\begin{tabular}{lcccS[table-format=2.2]}
\toprule
\textbf{Methodology} & \textbf{Top100 Builds} & \textbf{Awesome-CPP Builds} & \textbf{Total Builds} & \textbf{Success Rate (\%)} \\
\midrule
\textbf{Programmers}                                & 86 & N/A & N/A & N/A \\
\textbf{Default Build Commands}                     & 21 & 272 & 293 & 39.01 \\
\textbf{Bare LLM (\textit{DeepSeek-v2})}            & 23 & 239 & 262 & 34.22 \\
\updated{\textbf{Bare LLM (\textit{DeepSeek-v3})}}  & \updated{25} & \updated{264} & \updated{289} & \updated{38.43} \\
\textbf{Bare LLM (\textit{GPT-4o~mini})}            & 17 & 132 & 149 & 19.81 \\
\textbf{Bare LLM (\textit{GPT-4o})}                 & 23 & 215 & 238 & 31.65 \\
\updated{\textbf{\name (\textit{DeepSeek-v3})}}     & \updated{67} & \updated{510} & \updated{577} & \updated{76.73} \\
\textbf{\name (Default)}                            & \updated{\textbf{75}} & \updated{\textbf{512}} & \updated{\textbf{587}} & \updated{\textbf{78.10}} \\
\bottomrule
\end{tabular}
}

\end{table}

\subsection{Component Design Analysis (RQ4)}
\label{ablation}
\begin{figure*}[htbp]
\centerline{\includegraphics[width=0.75\textwidth]{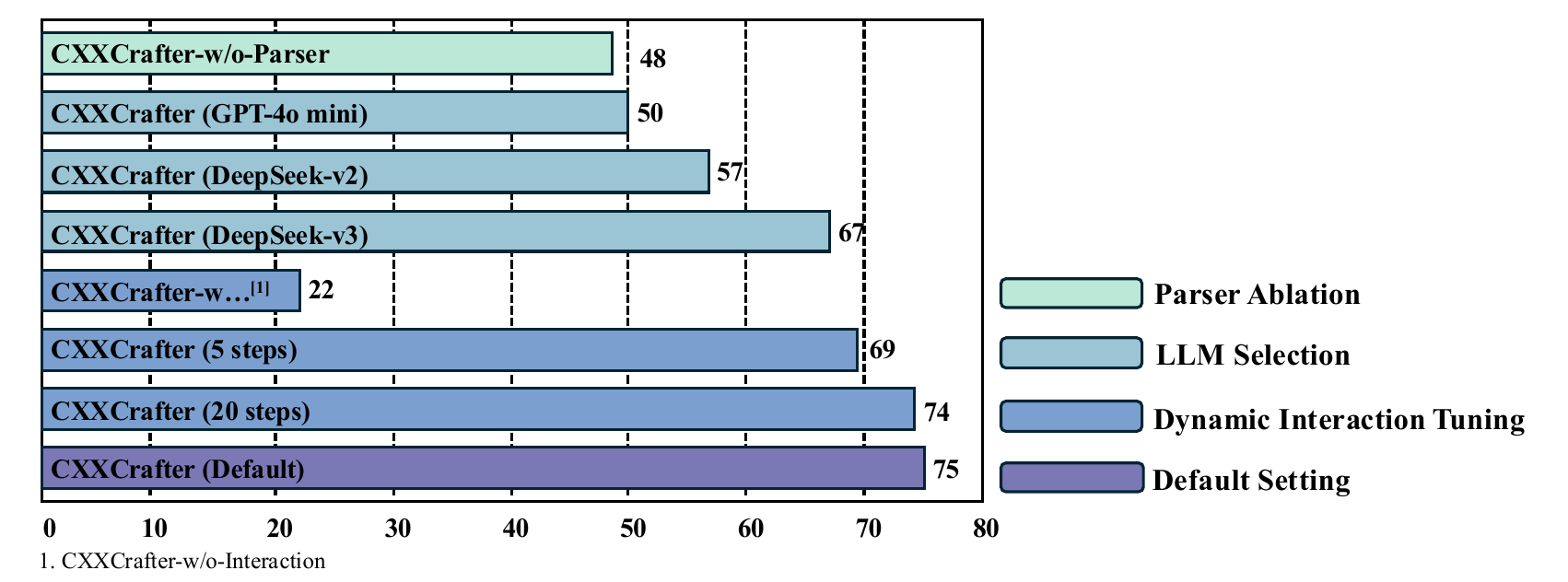}}
\caption{\updated{Number of Successful Builds of \name Variants on the \updated{Top100 Dataset.}}}
\label{fig: ablation}
\end{figure*}

In this section, we present a detailed component-wise analysis to assess the contribution of key modules and various configurations in \name. 
This analysis focuses on 3 main aspects:
\begin{itemize}
    \item The role of the \textbf{parser} module in enhancing build success.
    \item The impact of \textbf{dynamic interaction} and effect of varying dynamic interaction step counts.
    \item The impact of \textbf{different LLMs} on \name’s performance.
\end{itemize}
We conduct experiments on the Top100 dataset, with results shown in Figure \ref{fig: ablation}. \updated{The \name \texttt{(Default)} also uses \textit{GPT-4o} as the LLM, with a maximum of 10 dynamic interaction steps.}

\textbf{The Role of the Parser.}
The default configuration, with all components enabled, achieves the highest number of successful builds, completing 75 builds. When the parser is removed (\texttt{CXXCrafter-w/o-Parser}), the success rate drops to 48 builds, highlighting the parser's crucial role. In \name, build system selection and entry file identification rely on the parser, which forms the foundation for the entire build process and helps avoid many errors. Additionally, build-related documentation is crucial. The parser automates the search for and interpretation of these documents, further enhancing the build success rate.

\textbf{The Impact of Dynamic Interaction.}
\updated{
Dynamic interaction is the key design of \name, allowing iterative execution and modification during the build process. When dynamic interaction is disabled (\texttt{CXXCrafter-w/o-Interaction}), the number of successful builds drops sharply to 22, highlighting its importance in managing complex, multi-step build scenarios. We also analyze the impact of different interaction step limits. When the limit is set to 5 steps, performance declines, with only 69 successful builds. Increasing the step count to 20 does not further improve performance, with 74 successful builds. We observe that the benefits of increasing interaction steps begin to diminish beyond a certain threshold. For example, increasing the step count from 0 to 5 leads to a significant improvement of 47 successful builds. However, increasing it from 5 to 10 only adds 6 builds. Furthermore, increasing from 10 to 20 results in one fewer successful build. This variation is likely caused by the inherent instability in the LLM's output.
}

\begin{tcolorbox}[colback=gray!10, colframe=black, boxrule=0.5mm, boxsep=0mm, left=1mm, right=1mm, top=1mm, bottom=1mm]
\textbf{Finding 5:} Dynamic interaction plays a crucial role in managing multi-step tasks in the agent design. Increasing interaction steps improves success rates while the enhancement can be limited.
\end{tcolorbox}
\textbf{The Impact of Different LLMs.}
\updated{
Finally, we evaluate the impact of different LLMs on \name. Specifically, \textit{DeepSeek-v2} completes 57 builds, \textit{DeepSeek-v3} completes 67, while \textit{GPT-4o~mini} completes 50. \textit{GPT-4o} remains the most effective, with 75 successful builds. These results highlight the significant impact of LLMs on \name's ability to automate the build process. Notably, we observe that open-source LLMs can now achieve performance on par with leading closed-source models. Furthermore, cost-effective closed-source models like \textit{GPT-4o~mini} can achieve about 50\% of the effectiveness in our design. Additionally, \name, based on these models, performs much better as an agent than bare LLMs (see Section \ref{sec: overall}), further demonstrating that our design leads to a substantial improvement in performance.
}

\begin{tcolorbox}[colback=gray!10, colframe=black, boxrule=0.5mm, boxsep=0mm, left=1mm, right=1mm, top=1mm, bottom=1mm]
\textbf{Finding 6:} The selection of LLMs significantly affects the agent's performance. More powerful models, such as \textit{GPT-4o}, can offer stronger assistance and enhance the overall effectiveness.
\end{tcolorbox}

\subsection{Case Study (RQ5)}
\updated{Among the Top100 dataset, 72 projects are successfully built manually as well as by \name in Section \ref{sec: overall}.
Three projects succeed in \name but fail in manual builds, while 14 projects are successful manually but failed by \name.}

\updated{We analyze the 3 cases where manual builds failed.}
\updated{\textbf{Case 1}}: When building \textit{CQuery} manually, an error occurs with ``\textit{std::unique\_lock} being defined in the header \textit{<mutex>}'', suggesting that ``\textit{<mutex>}'' is not included. Initially, we suspect this is related to the \textit{Clang} version.
However, after trying various versions (e.g., \textit{Clang} 7, \textit{Clang} 11), the issue remains. Further analysis reveals that the \textit{CQuery} build script defaults to \textit{Clang} 7 on Ubuntu 14.04, which is incompatible with Ubuntu 22.04. \name suggests using Ubuntu 20.04, where \textit{Clang} 7 is  compatible and thus resolve the issue.
\updated{\textbf{Case 2}}: In the \textit{Paddle} project, human attempts produce the error: ``\textit{libdnnl.so}: undefined reference to \textit{dnnl::impl}...''. Despite confirming \textit{oneDNN} installations and testing various versions, the error persists. \name identifies a version mismatch between \textit{protobuf} and \textit{oneDNN} (requiring \textit{protobuf} 3.20.2+ as noted in the ``requirements.txt''), a detail overlooked by human builders.
\updated{\textbf{Case 3}}: In \textit{DOOM}, a linking error initially suggests issues with 32-bit libraries or the container environment, leading to various adjustments. \name identifies the actual issue as a mismatch between the \textit{TSL} variable \textit{errno} and the shared library version, resolving it with a code modification.

\updated{
\textbf{\name's Advantages over Humans.}
\name offers two key advantages over manual building: (i) The parser module uses RAG to efficiently process documents and other information, allowing it to identify build-related information more comprehensively than manual searches. For example, in CASE2, \name prevented errors that would have arisen from overlooking crucial information during manual builds. (ii) The LLM stores historical build knowledge, compensating for the limitations of human experience. As demonstrated in CASE3, \name makes more correct decisions, avoiding potential errors.
}
\updated{
The major drawback of \name is its higher error rate when installing complex dependencies, such as `CUDA' for \textit{OpenPose}. These libraries involve complex installation processes with many dependencies and steps. This may be resolved in the future through knowledge injection or RAG.
}

\subsection{Efficiency and Cost (RQ6)}
We assess the cost of \name across three dimensions: time, financial expense, and disk storage. \updated{These factors are crucial in determining the practical usefulness and scalability of our approach.}

\textbf{Time Cost.}
\updated{On average, \name takes 875.31 seconds to successfully build a project on the Top100 dataset. Additionally, the average time cost for the failed projects is 2.67 hours.} However, time costs can vary due to factors such as Docker caching and network speed. Enabling multi-processing significantly enhances efficiency, substantially reducing \updated{the overall build time}.

\textbf{Financial Cost.}
\updated{Running \name on the Top100 dataset generates 4,297,652 input tokens and 624,170 output tokens.}
This incurs \textit{GPT-4o} related costs of \$21.49 for input tokens and \$9.36 for output tokens, respectively. Among these, 75 projects \updated{are} successfully built, with an average cost per successful build calculated at \$0.41.
\updated{
The 25 failed projects generate a total of 2,420,092 input tokens and 225,154 output tokens, with an average cost of \$0.6191 per project. This price is based on OpenAI's pricing as of September 2024.
}

\textbf{Disk Storage Cost.}
The experiment generates over 50 TB of data, including Docker container caches and image files. This creates significant storage demands. Despite using three machines, disk space management remains a critical and recurring challenge throughout the experiment.
\section{Discussion}

\updated{
\textbf{Effectiveness and Consistency of Build Artifacts.}
\label{effectiveness_metrics}
We conduct an in-depth analysis of the build artifacts to verify their functionality and consistency with manually built artifact.
To verify that the build artifacts perform as expected, we run the unit tests provided by the projects. Among the 75 successfully built projects in TOP100, we identify 24 that generate test executables. Among them 22 projects pass while 2 fail due to missing audio connections in \textit{libsoundio} and lack of GUI display support for \textit{Stockfish} on our server (due to the absence of the relevant devices). These results confirm that the build artifacts produced by \name are valid and function as intended.
Additionally, we use a diff tool to compare the automated build artifacts with the manually built artifacts for all 75 successfully built projects. The results show that the automated and manual build artifacts are completely consistent.
Detailed experimental results can be found in our Project~\cite{zenodo_cxxcrafter}.
These results also further validate the effectiveness of our success metrics in the experiment.
}

 \updated{
\textbf{Building Different Software Versions.}
We conduct two additional experiments.
First, we investigate the build success rate across different software versions.
For 20 projects that are successfully built, we randomly select 5 commits for each, covering their entire repository commit histories from the creation of the repository.
\name achieves an 81\% success rate with 81 out of 100 builds successful.
This demonstrates that \name are effective across multiple versions. Some failures occur due to older versions requiring outdated packages, which are often hard to find.
Second, we test the build performance of consecutive commits (i.e., building one commit after another). 
By selecting the latest 5 commits from 20 projects, we observe a higher success rate, with 96 out of 100 builds successful.
Overall, these experiments demonstrate that \name is effective in both version diversity and consecutive commit builds.
}

 \updated{
\textbf{Building Different Language Projects.}
\name's design shows promising potential for other languages. Specifically, we conduct a simple migration to the top 100 most starred Java projects (after filtering the unbuildable projects, 76 remain), successfully building 57 out of 76 projects,
achieving a success rate of 75\%. This already represents promising performance in Java automation build methods \cite{hassan2017automatic} to our best knowledge.
}

\textbf{\updated{Potential} Applications of \name.}
\updated{
\name is highly beneficial for various downstream applications in software security analysis, including but not limited to: (1) reproducing identified vulnerabilities by facilitating set up environments with specific versions for vulnerability reproduction; (2) static program analysis, particularly high-precision analysis based on \textit{LLVM IR}, which often requires code to meet compilation requirements. \name can assist in fulfilling this process; (3) dynamic program analysis, such as source code instrumentation, where \name can ensure proper compilation, thus streamlining workflows for tasks like fuzz testing.
}
\updated{
\section{Threats to Validity}
Our study mainly suffers from the following threats to validity. 
Specifically, the internal validity threat in our study mainly stems from the variations in LLM performance, which could impact the experimental results. To mitigate this issue, we conduct experiments using the open-source model \textit{DeepSeek}. Additionally, during the dependency download process, \name retrieves dependencies from online sources (e.g., by using `\textit{apt}' to install packages or `\textit{git}' to download repositories). If these sources become unavailable or if network issues arise, it may affect the results. Furthermore, updates to the software itself could also introduce internal validity threats.
Our research primarily targets popular projects, for which LLMs may have gain deeper understanding and the documentation is usually more comprehensive. Therefore, \name's performance on less popular projects may be impacted, which constitutes an external validity threat. To address this, we plan to further incorporate RAG techniques or retrain the model in the future.
Lastly, the authors have rich experiences in C/C++ related researches, and via further extensive investigations in C/C++ projects build automation, the authors have gained deep understandings towards build-related issues. As a result, we believe this study' threats   of construct validity are limited.
}

\section{Conclusion}
This paper presents \name, an LLM-based agent system that automates the C/C++ build process, marking the first exploration of using LLM agents for this task. Through an empirical study of popular open-source C/C++ projects, we identify and categorize 384 build errors, providing insights into the key challenges of automating C/C++ builds. \name addresses these challenges by dynamically managing dependencies, resolving build issues, and diagnosing errors, achieving a success rate of 78\% across 752 projects. Our system advances build automation and offers significant support for downstream program analysis tasks, such as vulnerability research and performance optimization. Future work focuses on refining and expanding \texttt{CXXCrafter}’s capabilities to handle more complex build scenarios and support downstream program analysis tasks.

\section{Data Availability}
We release the implementation and all associated publicly available data in the website \cite{zenodo_cxxcrafter}.
\section*{Acknowledgment}
We would like to thank the anonymous reviewers for their insightful comments that helped improve the quality of the paper. 
This work was supported in part by the National Natural Science Foundation of China (U2436207, 62172105, 62372193). 
Yuan Zhang and Min Yang are the corresponding authors.
Yuan Zhang was supported in part by the Shanghai Pilot Program for Basic Research - FuDan University 21TQ1400100 (21TQ012).
Min Yang is a faculty of Shanghai Institute of Intelligent Electronics \& Systems, and Engineering Research Center of Cyber Security Auditing and Monitoring, Ministry of Education, China.
\bibliographystyle{ACM-Reference-Format}
\bibliography{References}

\end{document}